\documentclass[journal = aelccp, manuscript=article, layout=twocolumn]{achemso}

\usepackage{amsmath}
\usepackage{graphicx}
\usepackage{dcolumn}
\usepackage[english]{babel}

\usepackage{color}
\usepackage{float}

\usepackage{amssymb}

\let\oldmaketitle\maketitle
\let\maketitle\relax

\author{Sergei M. Butorin}
\email{sergei.butorin@physics.uu.se}
\affiliation{Condensed Matter Physics of Energy Materials, X-ray Photon Science, Department of Physics and Astronomy, Uppsala University, P.O. Box 516, SE-751 20 Uppsala, Sweden}

\title{Band gaps of hybrid metal halide perovskites: efficient estimation}

\begin{document}


\twocolumn[
\begin{@twocolumnfalse}
\oldmaketitle

\begin{abstract}
The employment of the parameter-free Armiento-K\"{u}mmel generalized gradient approximation (AK13-GGA) exchange functional was examined as means of the band gap prediction for hybrid metal halide perovskites (HaPs) or systems with strong spin-orbit coupling in the full-relativistic density-functional-theory (DFT) calculations. The new combination of AK13 with the nonseparable gradient approximation Minnesota correlation functional (GAM) was established as an approach allowing for the efficient band gap estimation with accuracy similar to the GW approximation method but at the computational costs of conventional DFT. This was further supported by results of the AK13/GAM calculations performed for various HaPs. The described approach creates an opportunity for the effective assessment of the electronic structure of large, complex, doped or defective HaPs and modelling of new materials.
\end{abstract}

\end{@twocolumnfalse}
]

\begin{tocentry}
\includegraphics[height=4.5cm]{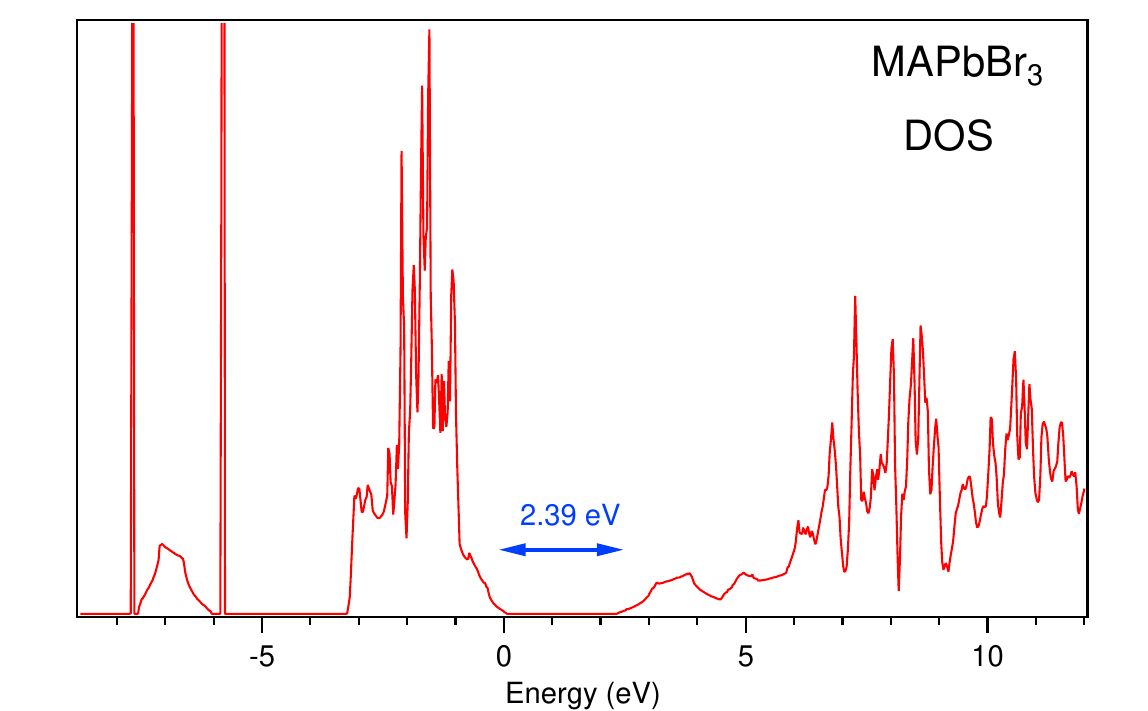}
\end{tocentry}



\section{Introduction}

Hybrid metal halide perovskites (HaPs) with Pb, Bi, Sn and Sb for metal are considered to be advanced materials for solar cell applications \cite{Kojima}. Related physics and chemistry depend on the size of the band gap and processes involving the states close to the valence band maximum (VBM) and conduction band minimum (CBM). However, conventional density functional theory (DFT) with usually good predictive ability fails to estimate correctly the size of the band gap in HaPs (Ref.\cite{Du}) when spin-orbit coupling (SOC) is taken into account in calculations. Since SOC is significant in HaPs, it can not be neglected. Furthermore, the role of SOC in the related processes needs to be studied. The use of hybrid functionals (e.g. Ref.\cite{Du}) and the GW approximation approach (e.g. Ref.\cite{Bokdam}) help to reproduce the right band gap sizes in HaPs but require extensive computational resources making calculations difficult for compounds with large number of atoms in the unit cell and other complex systems and interfaces. There is a need for less expensive methods with similar accuracy (for example, based on applications of suitable local density approximation (LDA) or generalized gradient approximation (GGA) functionals).

\begin{table*}
\centering
\caption{Shifted $k$-point mesh and optimized crystal structure data (adopted from references indicated in the table) used in the calculations of the electronic structure of HaPs. Orth stands for orthorombic, tetr stands for tetragonal, mon stands for monoclinic and all the other structures are pseudocubic.}
\begin{tabular}{lcc}
Compound&Shifted $k$-point mesh&Optimized crystal structure\\
\hline
MAPbCl$_3$&10x10x10&Ref.\cite{Walsh_collection}\\
FAPbCl$_3$&10x10x10&Ref.\cite{Bokdam} \\
CsPbCl$_3$-orth&8x8x10&Ref.\cite{Jocic}\\
MAPbBr$_3$&10x10x10&Ref.\cite{Walsh_collection}\\
FAPbBr$_3$&10x10x10&Ref.\cite{Bokdam}\\
CsPbBr$_3$-orth&8x8x6&Ref.\cite{Jocic}\\
MAPbI$_3$-tetr&6x6x6&Ref.\cite{Walsh_collection}\\
FAPbI$_3$&10x10x10&Ref.\cite{Walsh_collection}\\
CsPbI$_3$-orth&8x8x6&Ref.\cite{Jocic}\\
HdAPbI$_4$-mon&4x6x6&Ref.\cite{Safdari}\\
Cs$_2$AgBiCl$_6$&10x10x10&Ref.\cite{Walsh_collection}\\
Cs$_2$AgBiBr$_6$&10x10x10&Ref.\cite{Walsh_collection}\\
Cs$_2$AgBiI$_6$&10x10x10&Ref.\cite{Walsh_collection}\\
MASnCl$_3$&10x10x10&Ref.\cite{Bokdam}\\
MASnBr$_3$&10x10x10&Ref.\cite{Bokdam}\\
MASnI$_3$&10x10x10&Ref.\cite{Bokdam}\\
CsSnI$_3$-orth&8x6x8&Ref.\cite{Walsh_collection}\\
\end{tabular}
\label{table1}
\end{table*}

For HaPs, it was suggested to use the DFT-1/2 approach \cite{Tao} or to take advantage of the mBJ (Ref.\cite{Jishi}) and GLLB-SC (Ref.\cite{Castelli}) potentials for the calculation of the electronic structure and estimation of the band gaps. The calculated band gap sizes using these methods were reported to be in fair agreement with results of the GW calculations or/and experimentally determined values for a number of HaPs. Nevertheless, while minimizing the computational cost, these methods can bear some uncertainties.

In the DFT-1/2 approach \cite{Ferreira}, the parameter (cutoff radius $r_{cut}$) value needs to be set for each chemical element in the compound under study and that can cause alternative decisions on the choice of the orbital symmetry ($s$ or $p$ or $d$) for removing half an electron (or a quarter of an electron) from for the same element (see e.g. \cite{Tao,Traore}). Although, the unified optimized set of the $r_{cut}$ values is claimed to be transferrable from one HaP to another, in reality, these values require some re-adjustment for each compound if the criterium of obtaining the maximum band gap size is to be fulfilled. In case of the scheme using the mBJ potential \cite{Tran}, the re-adjustment of its parameters were also required in order to reproduce the correct sizes of the band gaps in HaPs (see e.g. \cite{Jishi,Traore2}). The calculations using the GLLB-SC potential \cite{Gritsenko,Kuisma} which were performed for HaPs (Ref.\cite{Castelli}) did not include SOC. The unified SOC correction to the GLLB-SC-calculated band gaps was made based on results from another DFT code \cite{Castelli}.

Another scheme which can be suggested to keep the computational cost similar to standard LDA and GGA and the accuracy comparable to the GW method in case of HaPs is to use the parameter-free AK13-GGA exchange functional by Armiento and K\"{u}mmel \cite{Armiento}. Our paper suggests the optimal combination of this exchange functional with a type of the correlation functional when applied to HaPs, presents the results of calculations of the band gap sizes for various HaPs and discusses some limitations of the AK13 functional in terms of the calculated electronic structure.


\section{Computational details}

\begin{table*}
\caption{Band gap size in MAPbBr$_3$ calculated at various levels of theory and compared with experimental value (in units of eV).}
\begin{tabular}{lcccccc}
Compound&AK13/NOC&AK13/WI&AK13/AM05&AK13/PW&AK13/GAM&Experiment\\
\hline
MAPbBr$_3$&1.89&1.93&1.98&1.99&2.39&2.30\\
\end{tabular}
\label{table2}
\end{table*}

To apply DFT, the Quantum Espresso v.6.8 code \cite{Giannozzi} was taken advantage of. The calculations were performed in the full-relativistic mode. First, combinations of the GGA exchange functional by Armiento and K\"{u}mmel (AK13) \cite{Armiento} (as it is defined in the LibXC v.5.1.6 library \cite{Lehtola}) with various correlation functionals were used to calculate the electronic structure of MAPbBr$_3$ (MA stands for methylammonium) and determine the size of the band gap. Then, a combination of AK13 with nonseparable-gradient-approximation Minnesota correlation functional by Yu \textit{et al.} \cite{Yu} (GAM) was applied in electronic-structure calculations for various Pb- and Bi-based HaPs. Furthermore, for Sn-based HaPs, a combination of AK13 with LDA correlation of the Perdew and Wang type \cite{Perdew_Wang} (PW) was also examined due to weaker SOC in those HaPs. The full-relativistic norm-conserving PBE (Perdew, Burke, and Ernzerhof \cite{Perdew}) pseudopotentials for lead, bismuth, tin, cesium, chlorine, bromine, iodine and silver were generated by the code of the ONCVPSP v.4.0.1 package \cite{Hamann} using input files from the SG15 database \cite{Scherpelz} (see Supplementary information (SI)). An additional feature of this ONCVPSP version is its ability to check for positive ghost states. For hydrogen, carbon and nitrogen, the full-relativistic norm-conserving PBE pseudopotentials in the UPF format from the SG15 database were used. The valence configurations for the pseudopotentials were defined as 1s$^1$ for H, 2s$^2$2p$^2$ for C, 2s$^2$2p$^3$ for N, 3s$^2$3p$^5$ for Cl, 4s$^2$4p$^5$ for Br, 4s$^2$4p$^6$4d$^9$5s$^2$ for Ag, 4d$^{10}$5s$^2$5p$^2$ for Sn, 4d$^{10}$5s$^2$5p$^5$ for I, 5s$^2$5p$^6$6s$^1$ for Cs, 5d$^{10}$6s$^2$6p$^2$ for Pb, and 5d$^{10}$6s$^2$6p$^3$ for Bi. The plane-wave cut-off energy was set to 60 Ry. The convergence threshold for density was 1.0x10$^{-12}$ Ry. The Van der Waals correction was applied using Grimme's D2 method \cite{Grimme}. The Brillouin zone was sampled using the Monkhorst-Pack scheme \cite{Monkhorst} and sizes of the $k$-point mesh for each compound are indicated in Table~\ref{table1} (where FA stands for formamidinium). The calculations were performed for the optimized crystal structures (using the conventional unit cell) adopted from Refs.\cite{Walsh_collection,Bokdam,Jocic} (see Table~\ref{table1}) where the structures were optimized using the PBEsol functional \cite{Perdew_PBEsol}. The structures optimized with this functional were used because it has been shown \cite{Lindmaa} that AK13 is not accurate for the geometry optimization. In the case of HdAPbI$_4$ (HdA stands for NH$_3$(CH$_2$)$_6$NH$_3$), the experimental crystal structure from Ref.\cite{Safdari} was used.

\section{Results and discussion}

\begin{figure}[h]
\includegraphics[width=\columnwidth]{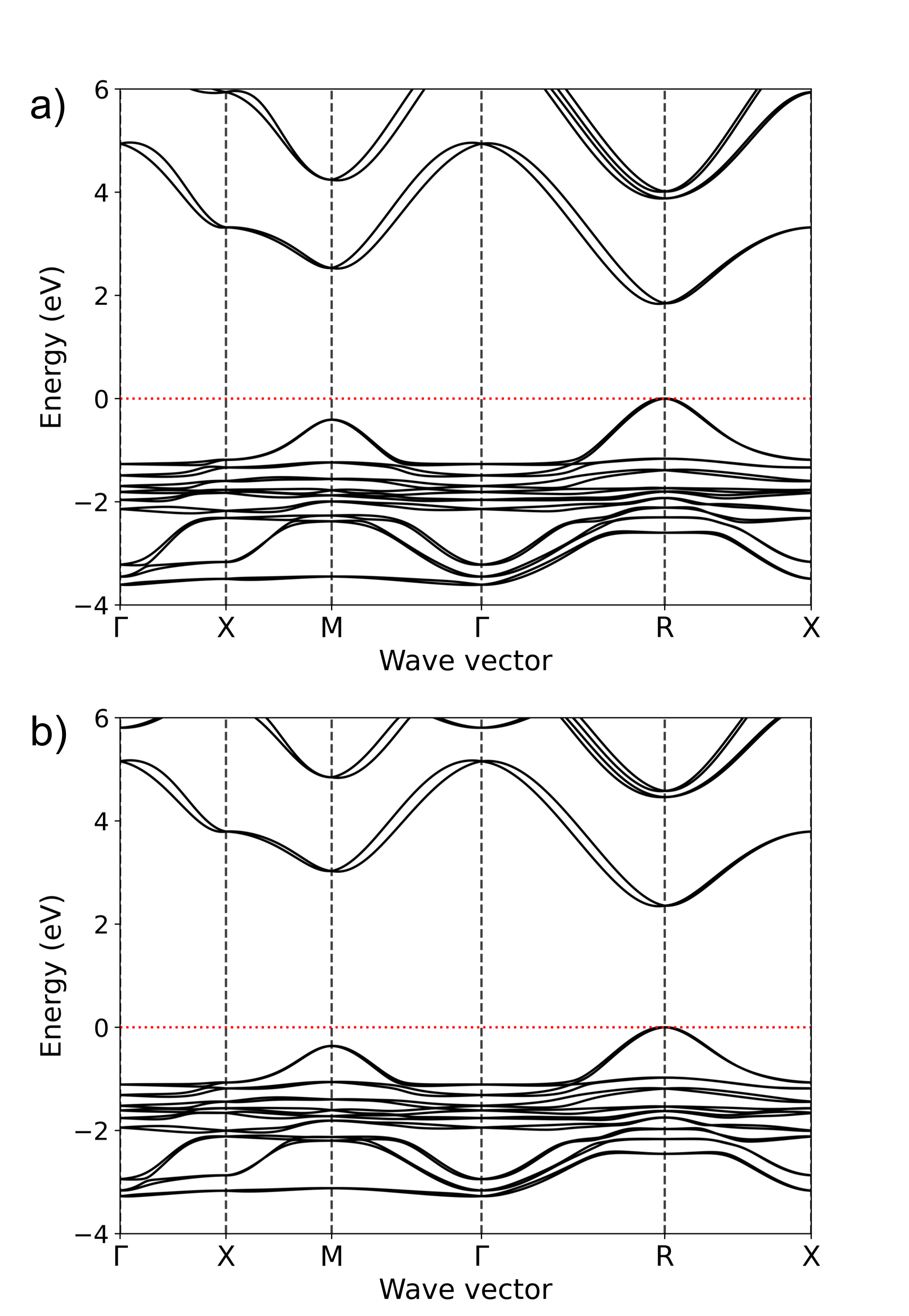}
\caption{Bandstructure of pseudocubic MAPbBr$_3$ calculated at the level of a) AK13/NOC and b) AK13/GAM theory. Zero eV is at the valence band maximum. \label{MAPbBr3_bands}}
\end{figure}

\begin{table*}
\centering
\caption{Calculated and measured band gaps of Pb- and Bi-based HaPs (in untis of eV).}
\begin{tabular}{lccc}
Compound&AK13/GAM&GW&Experiment\\
\hline
MAPbCl$_3$&2.90&3.07 (Ref.\cite{Bokdam})&2.94 (Ref.\cite{Baikie})\\
FAPbCl$_3$&2.88&3.07 (Ref.\cite{Bokdam})&3.01 (Ref.\cite{Wang_FAPbCl3}) \\
CsPbCl$_3$-orth&2.98&3.38 (Ref.\cite{Lang})&3.06 (Ref.\cite{Baranowski})\\
MAPbBr$_3$&2.39&2.34 (Ref.\cite{Bokdam})&2.22 (Ref.\cite{Baikie})\\
FAPbBr$_3$&2.40&2.26 (Ref.\cite{Bokdam})&2.23 (Ref.\cite{Eperon})\\
CsPbBr$_3$-orth&2.62&2.66 (Ref.\cite{Lang})&2.34 (Ref.\cite{Yang})\\
MAPbI$_3$-tetr&1.42&1.67 (Ref.\cite{Bokdam})&1.51 (Ref.\cite{Baikie})\\
FAPbI$_3$&1.16&1.48 (Ref.\cite{Bokdam})&1.48 (Ref.\cite{Eperon})\\
CsPbI$_3$-orth&1.54&1.81 (Ref.\cite{Lang})&1.72 (Ref.\cite{Yang})\\
HdAPbI$_4$-mon&2.30& &2.44 (Ref.\cite{Safdari})\\
Cs$_2$AgBiCl$_6$&2.45&2.42 (Ref.\cite{Filip})&2.77 (Ref.\cite{McClure})\\
Cs$_2$AgBiBr$_6$&2.18&1.83 (Ref.\cite{Filip})&2.19 (Ref.\cite{McClure})\\
Cs$_2$AgBiI$_6$&1.33& &1.75 (Ref.\cite{Creutz})\\
\end{tabular}
\label{table3}
\end{table*}

Initially, the AK13 functional was used without correlation in the DFT calculations \cite{Armiento,Vlcek,Tran2,Tran3}. Later, in some publications it was claimed that a combination of the AK13 with LDA (PW) correlation functional produces better results (see e.g. Refs.\cite{Cerqueira,Borlido}). Therefore, in present work, an attempt was made to find the right combination of AK13 with some correlation functional which would provide a correct result in terms of the band gap size in HaPs. Using MAPbBr$_3$ as a test compound, a number of combinations of AK13 with various correlation functionals from LibXC were applied to calculate the electronic structure and the band gap size in this compound. As a reference point, the result of applying AK13 with no correlation (NOC) was used. In this case, the calculated band gap of MAPbBr$_3$ was 1.89 eV which is smaller than the experimentally-established value of 2.30 eV. Table~\ref{table2} contains the results of only those combinations when the calculated band gap size is larger than that in the AK13/NOC case. Although, using AK13 with GGA correlation functionals by Wilson and Ivanov (WI) \cite{Wilson}, Armiento and Mattsson (AM05) \cite{Armiento2}, as well as with LDA (PW) correlation functional leads to increased band gaps, the calculated sizes are still smaller than the experimental value. A significant improvement in terms of agreement between calculated and measured values was obtained when AK13 was combined with GAM (see Table~\ref{table2} and Fig.~\ref{MAPbBr3_bands}). The results stands out and calculated 2.39 eV is quite close to the experimental observation.

Consequently, the AK13/GAM combination was further used in the DFT calculations to predict the band gap sizes in the Pb- and Bi-based HaPs. Table~\ref{table3} shows a comparison of the band gaps calculated at the AK13/GAM level of theory with experimental values for a number of HaPs. The published results obtained using the GW method are also included with corresponding references. For some compounds, the reported experimental data comprise a range of values for the band gap depending on whether the measurements were performed for policrystalline material in the form of powder or films or single crystals. In attempt to avoid a significant expansion of the reference list, a single, representative or average value of the measured band gap with the corresponding reference were chosen for the table.

Overall, the AK13/GAM-calculated band gaps of Pb- and Bi-based HaPs in Table~\ref{table3} are in similar agreement with experimental data as the GW band gaps. For example, for APbX$_3$ with X = Cl or I, the AK13/GAM band gaps are slightly underestimated while for X = Br they are slightly overestimated but the calculated values are relatively close to the measured ones. Similar predictive ability in terms of band gap estimations, as in case of the GW method, makes the AK13/GAM approach attractive for calculations for large, complex, doped or defective HaPs systems.

The AK13/GAM approach seems to work well for compounds with strong SOC but it may lead to a significant overestimation of the band gap size for compounds with relatively weak SOC. For latter compounds, combinations of AK13 with other correlation functionals may produce satisfactory results. For Sn-based HaPs, Table~\ref{table4} additionally contains the band gaps calculated with the AK13/PW combination. One can see that the differences between results obtained at the AK13/GAM and AK13/PW levels of theory become less pronounced when compared to the case of Pb-based HaPs.

\begin{table*}
\centering
\caption{Calculated and measured band gaps of Sn-based HaPs (in untis of eV). Note that the band gap in MASnCl$_3$ was measured for the monoclinic phase.}
\begin{tabular}{lcccc}
Compound&AK13/GAM&AK13/PW&GW&Experiment\\
\hline
MASnCl$_3$&3.57&3.47&4.02 (Ref.\cite{Bokdam})&3.61 (Ref.\cite{Wang})\\
MASnBr$_3$&2.44&2.07&1.87 (Ref.\cite{Bokdam})&2.15 (Ref.\cite{Hao})\\
MASnI$_3$&1.20&1.08&1.03 (Ref.\cite{Bokdam})&1.21 (Ref.\cite{Stoumpos})\\
CsSnI$_3$-orth&1.28&1.16&1.34 (Ref.\cite{Lang})&1.30 (Ref.\cite{Stoumpos})\\
\end{tabular}
\label{table4}
\end{table*}

\begin{figure}[h]
\includegraphics[width=\columnwidth]{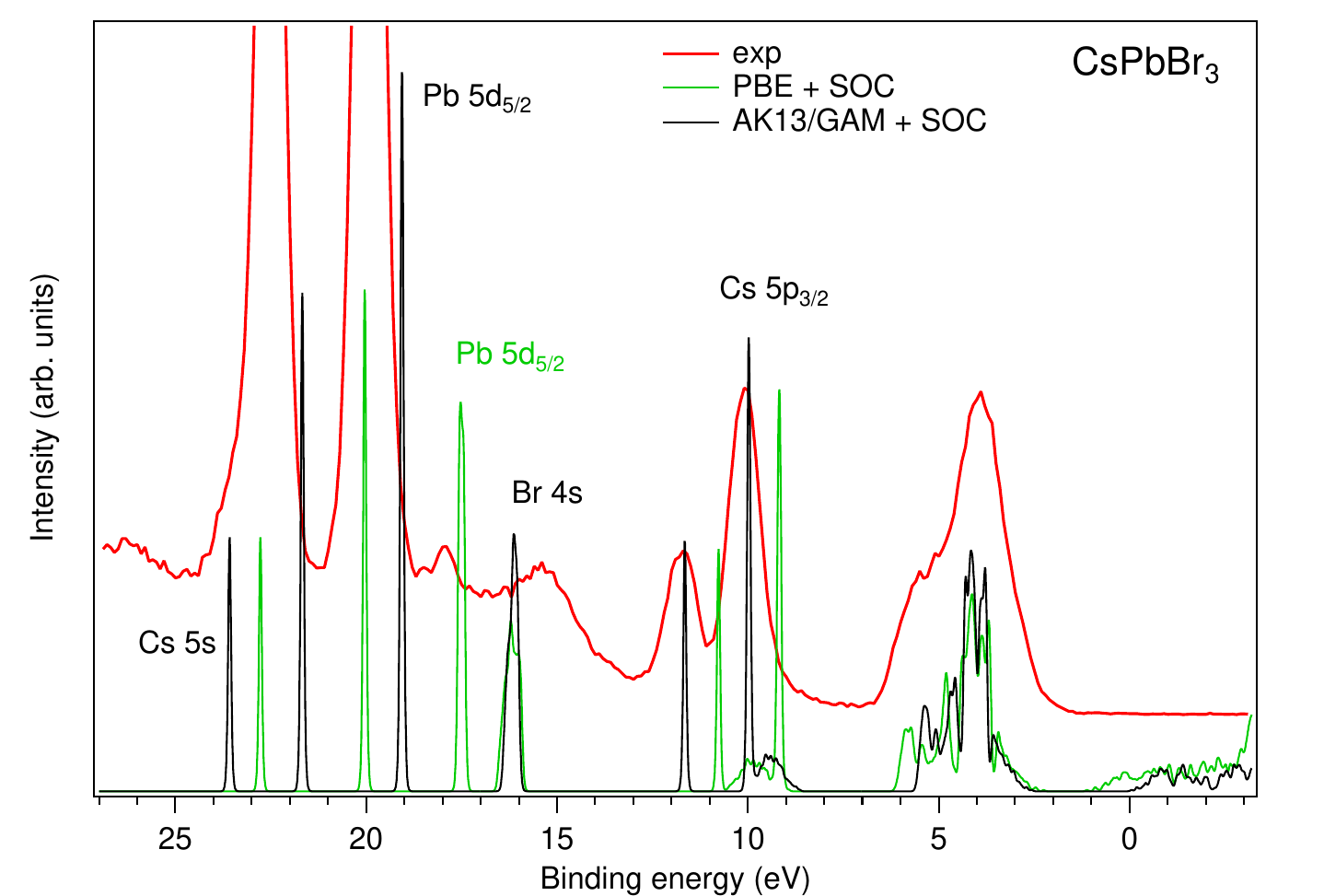}
\caption{Comparison of total density of states of CsPbBr$_3$ calculated using AK13/GAM (black curve) and PBE (green curve) functionals with the HAXPES spectrum (red curve) published in Ref.\cite{Man2}. The calculated DOS curves are aligned with the HAXPES spectrum using the intensity maximum of the valence band. \label{XPS_CsPbBr3}}
\end{figure}

Another important finding when using the AK13/GAM approach for HaPs is that the energy positions of the shallow core levels are more accurately calculated as compared to the results of the PBE calculations commonly used to describe x-ray photoemission (XPS) spectra of HaPs (e.g. Refs.\cite{Phuyal,Phuyal2,Man,Man2,Sterling}) which probe the occupied density of states (DOS). Fig.~\ref{XPS_CsPbBr3} compares the total DOS of CsPbBr$_3$ calculated using the AK13/GAM and PBE functionals with hard x-ray photoemission (HAXPES) spectrum of CsPbBr$_3$ recorded at the incident photon energy of 4000 eV. The total DOS was broadened by Gaussian with full width at half maximum (FWHM) of 25 meV. CsPbBr$_3$ was chosen as an example because its XPS/HAXPES spectrum contains, besides Pb $5d_{5/2}$ and $5d_{3/2}$ lines, the Cs $5p_{3/2}$ and $3p_{1/2}$ doublet close to the valence band.

An inspection of Fig.~\ref{XPS_CsPbBr3} reveals that in addition to the more correct description of the size of the band gap by the AK13/GAM calculations as compared to the PBE case, the energy positions of the core levels appear to be a better match for experimental ones than in the PBE case. The AK13/GAM-calculated energies of the Cs $5p_{3/2,1/2}$ levels are in good agreement with experiment while the offset of the AK13/GAM-calculated Pb $5d_{5/2,3/2}$ levels with respect to measured ones is much less than in the PBE calculations. It is interesting that the comparison of the AK13/GAM-calculated total DOS with the HAXPES spectrum suggests the Fermi level (chemical potential for electrons) position to be close to CBM, thus indicating rather the $n$-type semiconductor character for CsPbBr$_3$.

Fig.~\ref{XPS_CsPbBr3} also shows that the width of the valence band calculated by the AK13/GAM method appears to be smaller as compared to the PBE calculations. As it was previously pointed out in Refs.\cite{Vlcek,Tran3}, AK13 somewhat underestimates the band dispersion, which in turn can lead to an overestimation of the effective hole and electron masses. Another discussed limitation \cite{Tran4} of AK13 is a difficulty in calculations for atomically thin films often referred to as two-dimensional (2D) systems, because AK13 leads to numerical problems due to the presence of vacuum. However, there was now problem to perform the AK13/GAM calculations and obtain the band gap estimation for the layered 2D material, such as HdAPbI$_4$.

\section{Conclusions}

The overall results reported here show a very useful utility of the AK13 functional for the prediction of the band gaps in HaPs. The AK13/GAM combination allows one to keep the computational cost at the level of standard DFT while providing the accuracy similar to that of the GW method. In particular, the employment of the described approach can be taken advantage of in calculations of the electronic structure of large, defective, doped or more sophisticated, newly modelled HaPs systems.

Furthermore, a comparison with experimental data, which probe the occupied states, shows that the use of the AK13 functional leads to a better description of the shallow core levels in the calculated DOS of HaPs as compared to rather common PBE approach. It is an important finding since the electrons from the shallow core levels are participating in the chemical bonding.

\textbf{Notes}
The author declare no competing financial interest.

\begin{acknowledgement}
The author acknowledges the support from the Swedish Research Council (research grant 2018-05525). The computations and data handling were enabled by resources provided by the Swedish National Infrastructure for Computing (SNIC) at National Supercomputer Centre at Link\"{o}ping University partially funded by the Swedish Research Council through grant agreement no. 2018-05973.
\end{acknowledgement}

\begin{suppinfo}
Input files to generate the full relativistic norm-conserving pseudopotentials using ONCVPSP v.4.0.1.
\end{suppinfo}
\bibliography{AK13}

\end{document}